\documentclass[%
reprint,
showkeys,
nofootinbib,
nobibnotes,
 amsmath,amssymb,
 aps,
pra,
 longbibliography,
 lengthcheck,%
]{revtex4-1}

\usepackage{graphicx}
\usepackage{dcolumn}
\usepackage{bm}
\usepackage{hyperref}
\usepackage{multirow} 
\usepackage{relsize}

\newcommand{\ar}{\rightarrow}

\newcommand{\kF}{k_{\rm F}}

\newcommand{\mi}{m_{\rm i}}
\newcommand{\mh}{m_{h}}

\newcommand{\<}{\mathsmaller{<}}
\renewcommand{\>}{\mathsmaller{>}}

\newcommand{\be}{\begin{equation}}
\newcommand{\ee}{\end{equation}}
\def\bea{\begin{eqnarray}}
\def\eea{\end{eqnarray}}
\newcommand{\w}{{\mathsf w}}

\usepackage{amsthm}

\begin{document}


\title{Kinetic theory for a mobile impurity in a degenerate Tonks-Girardeau gas}

\author{O. Gamayun$^{1,2}$}
\author{O. Lychkovskiy$^{1,3}$}
\author{V. Cheianov$^1$}
\affiliation{%
$^1$Lancaster University, Physics Department, Lancaster LA1 4YB, UK,
}%
\affiliation{$^2$Bogolyubov Institute for Theoretical Physics, 14-b Metrolohichna str., Kyiv 03680, Ukraine,}
\affiliation{$^3$Russian Quantum Center, Novaya St. 100A, Skolkovo, Moscow Region, 143025, Russia.}
%


\begin{abstract}

A kinetic theory describing the motion of an impurity particle in a degenerate Tonks-Girardeau gas is presented. The theory is based on the one-dimensional Boltzmann equation. An iterative procedure for solving this equation is proposed, leading to the exact solution in number of special cases and to an approximate solution with the explicitly specified precision in a general case. Previously we have reported that the impurity reaches a non-thermal steady state, characterized by an impurity momentum  $p_\infty$ depending on its initial momentum $p_0$ \cite{Burovski2013}. In the present paper the detailed derivation of $p_\infty(p_0)$ is provided. We also study the motion of an impurity under the action of a constant force $F$. It is  demonstrated that if the impurity is heavier than the host particles, $\mi>\mh$, damped oscillations of the impurity momentum develop, while in the opposite case, $\mi<\mh$, oscillations are absent. The steady state momentum as a function of the applied force is determined. In the limit of weak force it is found to be force independent for a light impurity and proportional to $\sqrt{F}$ for a heavy impurity.

\end{abstract}

\maketitle


\date{\today}

\section{\label{sec intro} Introduction}

Last decade has been witnessing a tremendous experimental progress in the fabrication and manipulation of artificial one-dimensional (1D) quantum systems, especially 1D ultracold atomic gases \cite{bloch2008many}. An important recent advancement is the development of the experimental techniques enabling injection and tracking of a single impurity in a 1D host gas of bosons. This opens access to previously unexplored aspects of non-equilibrium dynamics in 1D \cite{palzer2009quantum,weitenberg2011single,*catani2012quantum,*spethmann2012dynamics,*fukuhara2013quantum,*fukuhara2013microscopic}.

If prior to these experimental
developments theoretical studies of mobile impurities in 1D systems were mainly concerned with equilibrium properties and spectral characteristics (see \cite{imambekov2012one} and references therein), the theoretical focus has now shifted to the analysis of non-equilibrium time-resolved phenomena. A rich variety of such phenomena relating to the time evolution of the impurity momentum have recently been predicted.
These include quasi-Bloch oscillations \cite{Gangardt2009,schecter2012dynamics,Gangardt2012},
quantum flutter \cite{mathy2012quantum,knap2013quantum} and relaxation to a non-thermal state \cite{Burovski2013,Lychkovskiy2013,Gamayun2014keldysh,mathy2012quantum,knap2013quantum,Lychkovskiy2013}.


The present theoretical understanding of these phenomena is not complete and number of outstanding questions remain. For example, what are the physical conditions for the occurrence of oscillations? Are those oscillations damped at zero temperature and if so what factors determine damping rate? What are the properties of the non-thermal state? How does it depend on initial conditions?

In our earlier paper \cite{Burovski2013} we proposed to address these questions by considering a simple yet nontrivial model of a single impurity weakly coupled by a point-like interaction to a degenerate Tonks-Girardeau gas. Ref. \cite{Burovski2013} advocated the use of Boltzmann kinetic theory and
reported certain rigorous results on the properties of a non-thermal steady state in the absence of an external force. Applicability of the Boltzmann theory to the model under consideration was rigorously justified by means of Keldysh technique  in Ref. \cite{Gamayun2014keldysh}. In the present paper we give a more detailed derivation of the results of Ref. \cite{Burovski2013} and  analyze the dynamics of an impurity under the action of a constant force $F$.

Our main results are as follows. We demonstrate that the dynamics of the impurity under the action of a force depends dramatically on the impurity-to-host mass ratio. In the heavy impurity case oscillations develop while in the light impurity case the momentum of the impurity saturates without oscillations. The oscillations are shown to be damped with a rate proportional to $F^2$ in the limit of weak force. At large times a steady state is established. We determine the steady state momentum as a function of the applied force. In particular, in the limit of weak force the steady state momentum is found to be force independent for a light impurity and proportional to $\sqrt{F}$ for a heavy impurity.

The paper is organized as follows.  In Sec. \ref{Litra}  we describe the model, discuss various regimes and overview known results. In Sec. \ref{sec setup} we introduce basic notations and state precisely the problem to be solved. Sec. \ref{sec kinematics} is devoted to the peculiar one-dimensional kinematics of the problem. In Sec. \ref{sec Boltzmann equation} the Boltzmann equation without a force is introduced and an algorithm for its iterative solution is presented, along with exact solutions in special cases. In Sec. \ref{sec asymptotic distribution} the integral equation on the infinite-time momentum of the impurity (in the absence of a force) is introduced and an algorithm for its iterative solution is presented, along with exact solutions in special cases. Sec. \ref{sec force} is devoted to the dynamics of an impurity under the action of a constant force. The dynamics of the impurity and the steady state are quantitatively described. The summary and concluding remarks are given in Sec. \ref{sec summary and outlook}. In the Appendix the integral equation for the infinite-time momentum is derived from the Boltzmann equation.

\section{Model and overview of known results}
\label{Litra}

In this section we describe a model of an impurity immersed a 1D host gas and give a brief overview of recent results on the dynamics of the impurity at zero temperature. The host gas is a gas of bosons described by the Hamiltonian
\be
H_h = \sum_{n=1}^N \frac{p_n^2}{2m_h} + \gamma_h  \frac{\rho}{\mh}\sum_{n<n'}\delta(x_{n}-x_{n'})\,,
\ee
where  $p_n$ and $x_{n}$ are momentum and coordinate of the $n$'th host particle respectively, $\gamma_h$ is the dimensionless coupling constant
and $\rho\equiv N/L$ is the linear density of the host gas.
The impurity has mass $m_i$ and interacts with the host through the interaction term
\be
V  =  \gamma \frac{\rho}{\mh} \sum\limits_{n=1}^N \delta(X-x_n).
\ee
Here $X$ is coordinate of the impurity and interaction strength is characterized by the coupling constant $\gamma$. Here and in the following subscript $h$ and $i$ refer to the host and to the impurity, respectively.

The dynamics of the impurity is determined by the two
dimensionless coupling constants $\gamma$ and $\gamma_{h}$, by the mass ratio
$\eta \equiv m_i/m_h$ and by the constant force $F$ if it is applied to the impurity.

The value of the host-host coupling, $\gamma_{h}$, determines the physics of the host gas.  Two limiting cases are well understood theoretically. In the Bogolyubov limit of small $\gamma_{h}$ the problem is amenable to mean-field treatment \cite{Bogolyubov1947JPhys,Bogolyubov1947BulMSU}, while in the Tonks-Girardeau limit of infinitely large $\gamma_{h}$ the bosons can be mapped on the noninteracting fermions~\cite{Girardeau1960}.
From the experimental point of view, a wide range of values of $\gamma_{h}$ (as well as $\gamma$) can be explored due to the tunability of the scattering length through the Feschbach resonance \cite{bloch2008many}.

The mass ratio $\eta$ is on the order of unity in the case of atomic impurities studied in recent experiments
\cite{palzer2009quantum,weitenberg2011single,*catani2012quantum,*fukuhara2013quantum,*fukuhara2013microscopic}.
For the theoretical treatment the exact value of $\eta$ is of crucial importance. Firstly, if $\eta=1$  and either $\gamma_{h}=\infty$ \cite{mcguire1965interacting} or $\gamma_{h}=\gamma$ \cite{gaudin1967systeme,yang1967some}, the model is integrable by Bethe Ansatz.
Secondly, in the case of an impurity weakly coupled to a Tonks-Girardeau gas, $\eta=1$ is a singular point no matter whether or not integrability is present \cite{Burovski2013,Gamayun2014keldysh,Lychkovskiy2014}.
Finally, cases of light ($\eta <1$) and heavy ($\eta >1$) impurities qualitatively differ from each other and from the case of  equal masses \cite{Burovski2013,schecter2012dynamics,Gamayun2014keldysh}.



Theoretical papers \cite{Gangardt2009,schecter2012dynamics,Gangardt2012,mathy2012quantum,knap2013quantum,Burovski2013,Lychkovskiy2013,Gamayun2014keldysh} explored various regions of parameter space with a range of methods. In \cite{Gangardt2009,schecter2012dynamics,Gangardt2012} the motion of an impurity driven by a constant force was investigated by methods of quantum hydrodynamics. It was predicted that under certain conditions the momentum of an impurity immersed in a 1D gas exhibits oscillations resembling the Bloch oscillations in an ideal crystal~\cite{Gangardt2009}. A wide range of parameters was discussed in \cite{Gangardt2009,schecter2012dynamics,Gangardt2012}, however most of the quantitative results were obtained for a host gas in the Bogolyubov regime (small $\gamma_h$) or for the case $\gamma,\gamma_h \gg 1$.

The relaxation of the momentum of an impurity injected in a 1D gas with some initial momentum $p_0$ was studied in \cite{mathy2012quantum,knap2013quantum}. In Ref. \cite{mathy2012quantum} the integrable case $\eta=1$, $\gamma_{h}=\infty$ was mainly considered using a combination of analytical and numerical techniques. The nonintegrable case of nonequal masses and $4\leq\gamma_{h} \leq \infty$ was numerically studied in \cite{knap2013quantum}.  The impurity-host coupling in the range $3\leq\gamma \leq 20$ was considered in both Refs. \cite{mathy2012quantum,knap2013quantum}. Two intriguing effects were discovered in \cite{mathy2012quantum,knap2013quantum}.
Firstly, the impurity momentum as a function of time was found to exhibit oscillations around some average value. This phenomenon was dubbed ``quantum flutter'' \cite{mathy2012quantum}. Secondly, despite the observed damping of the oscillations the average momentum did not appear to change significantly with time.
This observation indicates that the momentum of the impurity at infinite time, $p_\infty$, is nonzero.  This could seem to be in apparent conflict with the predicted absence of superfluidity in the system \cite{Astrakharchik2004}, but in fact the conflict is absent \cite{Lychkovskiy2013}.

Recently we have reported analytical results on the relaxation of the impurity momentum \cite{Burovski2013} obtained for an impurity weakly coupled to a 1D Bose gas in the Tonks-Girardeau limit. In the general case of nonequal masses the reported results have been based on the Boltzmann equation.
The applicability conditions of the Boltzmann equation in 1D systems are subtle. For our system these include $\gamma^2\ll1$ and $|\eta-1|\gg\gamma$ \cite{Burovski2013}. A detailed derivation of the Boltzmann equation from the macroscopic quantum theory has been presented in the separate publication~\cite{Gamayun2014keldysh}. Previously the Boltzmann equation was applied to calculate the mobility of an impurity in a 1D gas at a finite temperature \cite{CastroNeto1995}.



\section{\label{sec setup}The scope of the present work}

\begin{figure*}[t]
\begin{tabular}{cc}
\includegraphics[width=\linewidth]{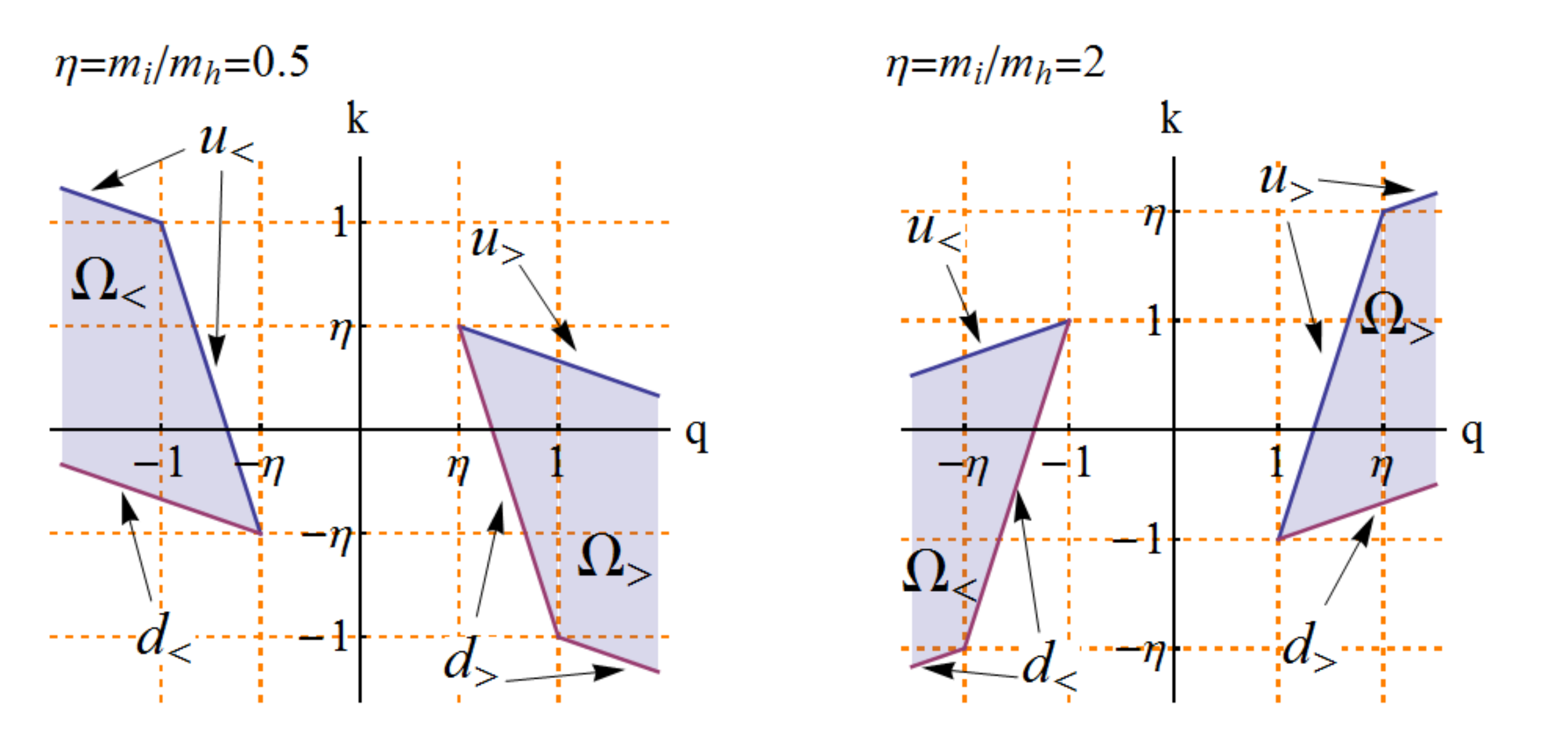}
\end{tabular}
\caption{
\label{fig kinematics}
Kinematically allowed regions for a single pairwise scattering in the cases of the light
(left) and heavy (right) impurity. The final momentum of the  impurity, $k$,   is shown {\it vs} the initial
momentum of the  impurity, $q.$}
\end{figure*}


We consider the Tonks-Girardeau limit: $\gamma_h=+\infty$. In this limit a gas of host bosons can be mapped on the gas of noninteracting fermions \cite{girardeau1960relationship}. We find that the fermionic description provides better insight into the physics of the problem. Therefore, we use this description in what follows. Needless to say, our results are equally applicable for a genuine free fermion host. The latter, however, is less relevant in the context of current  experiments \cite{bloch2008many}.

Thus, we consider a single impurity particle immersed in a gas of $N$ identical noninteracting fermions with density $\rho$. The Fermi momentum and time are defined as $\kF = \pi \rho$ and $t_{\rm F} \equiv 2 \mh/ \kF^2$, respectively. In order to simplify the notations we henceforth set $\kF =1$ unless explicitly specified. We work in the limit of weak impurity-host coupling $\gamma$.

Our main goal is to find the distribution function of the impurity momentum
$w_k(t)$ and investigate its behavior at $t\to\infty$. This can be used for the calculation of the average momentum of the impurity in the infinite time limit:
\be
p_\infty =\lim\limits_{t\to\infty} \sum_k k~ w_k(t)\,.
\ee
We assume the following initial conditions at $t=0$. The fermions are in the ground state of noninteracting Hamiltonian $H_h$ (Fermi sea). The impurity is in a statistical mixture of the impurity momentum eigenstates. Due to the linearity of the problem one can focus on the initial condition with a fixed impurity momentum,
\be\label{initial condition}
w_k(0)=\delta_{k,p_0},
\ee
where, without loss of generality,  $p_0>0$.
Where this specific form of the initial condition is essential, we use the notations $w_{p_0\to k}(t)$ and $w_{p_0\to k}^\infty$ instead of $w_k(t)$ and $w_k(t=\infty)$.

\section{\label{sec kinematics}Kinematics}

Understanding the kinematics of pairwise scattering in one dimension is essential for studying the dynamics of the impurity. From the classical point of view, a scattering event in one dimension  can be characterized by two momenta, e.g. the initial, $q$, and the final, $k$, momenta of the impurity. The other two (in our case, the initial $s$ and the final $l$ momenta of the host particle) are fixed by the energy and momentum conservation laws:
\be
\frac{q^2}{2m_i}+\frac{s^2}{2m_h} = \frac{k^2}{2m_i}+\frac{l^2}{2m_h}\,,
\ee
\be
q+s=k+l\,.
\ee
In addition, the Pauli principle applied to host fermions, imposes
\be
|s|<1\,\,\,,|l|>1\,
\ee
that restricts possible values of $k$ and  $q$  to a certain region $\Omega$ in the $(q,k)$ plane.

The region $\Omega$ is shown in Fig.~\ref{fig kinematics}. It consists of two disjoint pieces $\Omega = \Omega_\> \sqcup \, \Omega_\<$. We denote the upper and lower boundaries of $\Omega$ as $u(q)$ and $d(q)$, respectively. They also consist of disjoint pieces, $u(q)=u_\>(q)\sqcup u_\<(q)$ and $d(q)=d_\>(q)\sqcup d_\<(q)$, and are constructed from segments of the following lines:
\be\label{boundaries}
\left\{
\begin{array}{ll}
k_{\rm top}(q)      &= \frac{\eta -1}{\eta +1} q +  \frac{2\eta}{\eta +1}, \\
k_{\rm bot}(q)   & = \frac{\eta -1}{\eta +1} q -  \frac{2\eta}{\eta +1}, \\
k_{\rm left}(q)     & = \frac{\eta +1}{\eta -1} q +  \frac{2\eta}{\eta -1}, \\
k_{\rm right}(q)    & = \frac{\eta +1}{\eta -1} q - \frac{2\eta}{\eta -1}. \\
\end{array}
\right.
\ee

Notice that a segment of $k_{\rm left}(q)$  is a part of either the $u_\<(q)$, or  $d_\<(q)$, depending on the mass ratio $\eta$; an analogous statement holds for $k_{\rm right}(q)$.

Considering the initial momentum $p_0$ one can identify regions in the momentum space corresponding to the qualitatively different behavior of the impurity. For $|p_0|<q_0\equiv \min\{1,\eta\}$ the impurity can not scatter on host fermions due to the  Pauli blocking, so the impurity momentum is conserved. Obviously, if after several scattering events the impurity momentum drops below $q_0,$ scattering stops. Further, there exist a point $q_1>q_0$ such that, whenever $q_0<|p_0|<q_1$, a single scattering transfers the impurity momentum below $q_0$. In general, there exists an infinite ascending sequence $\{q_n\}$ with the following property: the impurity momentum is transferred below $q_0$ in no more than $n$ scattering events whenever $|p_0|<q_n.$ The recursive definition for the sequence reads
\be
q_{n-1} =\max\{|u(q_n)|,|d(q_n)|\}.
\ee
This sequence converges to $q_\infty=\max\{1,\eta \}$ with \mbox{$n\rightarrow\infty.$} The first three $q_n$ are given in Table \ref{table qn}.

\begin{table}[h]
\caption{\label{table qn}%
$q_0$, $q_1$, $q_2$ and $q_\infty$ for light and heavy impurity.
}
\begin{ruledtabular}
\begin{tabular}{l|ccccc}
& $q_0 $ & $q_1 $ & $q_2 $ & ... & $q_\infty $\\
\hline
$\eta<1$ & $\eta$ & $\frac{(3-\eta)\eta}{1+\eta}$ & $\frac{(\eta^2-2\eta+5)\eta}{(1+\eta)^2}$ & ... & 1\\
$\eta>1$ & 1 & $\frac{3\eta-1}{1+\eta}$ & $\frac{5\eta^2-2\eta+1}{(1+\eta)^2}$ & ... & $\eta$\\
\end{tabular}
\end{ruledtabular}
\end{table}

Note that the case of equal masses ($\eta=1$) is special: $\Omega_\>$ and $\Omega_\<$ are two rectangular horizontal stripes and the whole sequence $\{q_n\}$ collapses to a single point, $q_n=1 ~\forall n$. At first glance this may look as a simplification; however, in fact the kinematics of the equal mass case represents a major difficulty for derivation and application of the Boltzmann equation \cite{Burovski2013,Gamayun2014keldysh}, which is discussed in more detail below.

\section{\label{sec Boltzmann equation}Boltzmann equation}
\subsection{\label{subsec derivation of Boltzmann equation}Preliminary considerations}

\begin{table*}
\caption{\label{tab width}Total dimensionless decay width ${\bf \Gamma}_k$ (defined in Eq. \eqref{total dimensionless width}) calculated in the thermodynamic limit.}
\begin{ruledtabular}
\begin{tabular}{c|cccc}
 & $|k|<q_0$   &\multicolumn{2}{c}{$q_0 \leq |k| < q_\infty$} & $|k| \geq q_\infty$ \\
 &             & $\eta<1$ & $\eta>1$                          &                     \\
 \hline
${\bf \Gamma}_k$ &
0 &
$\log\frac{1+\eta}{1-\eta}$&
$\log\frac{\eta +|k|}{\eta - |k|}-\log\frac{\eta+1}{\eta-1}$&
$\log\frac{|k|+\eta }{|k|-\eta }$
\end{tabular}
\end{ruledtabular}
\end{table*}

The above kinematical reasoning can be used as a starting point for the Boltzmann kinetic theory. It is well known, however, that in one dimension this should be done with caution. Indeed, the validity of the Boltzmann equation relies on the Lorentzian
shape of the particle's spectral function such as in the Fermi liquid theory
\cite{Mahan2000}. At the same time spectral functions of particles in 1D  quantum fluids tend to exhibit essentially non-Lorentzian shapes in the vicinity of the mass shell \cite{Giamarchi2003,imambekov2012one}. This is not true, however,
when the mass shell of a particle falls into a broad
spectral continuum of excitations into which it can decay
\cite{khodas2007fermi}: in this case the quasi-Lorentzian shape of a  spectral function is restored. The latter fact makes it possible to apply Boltzmann's kinetic theory to our problem. The rigorous proof of this statement (based on resummation of relevant diagrams in Keldysh technique) is given in a separate publication~\cite{Gamayun2014keldysh}. Here, instead, we present a heuristic derivation of the Boltzmann equation based on the straightforward application of the leading order perturbation theory.
This approach quickly leads to a correct result and, as a side product, yields $p_{\infty}$ in the next-to-leading order ($\sim \gamma^2$) in the case when $|p_0|$ is below and sufficiently far from $q_0$ (see Eq. \eqref{pinfty with quantum corrections} below).
Afterwards, we summarize the applicability conditions  for  the Boltzmann equation which follow from the rigorous treatment \cite{Gamayun2014keldysh}.

The leading order of the time-dependent perturbation theory gives the following expression for $w_k(t)$:
\be\label{naive}
\frac{w_k(t)-w_k(0)}{t}=
 -  \tilde{\Gamma}_k(t) w_k(0)\\
+ \sum_q \tilde{\Gamma}_{q\to k}(t) w_{q}(0)\,.
\ee
Here
\be
\tilde{\Gamma}_{q\to k}(t) \equiv \gamma^2 \left(\frac{1}{\pi \mh L}\right)^2
\mathop{\sum_{|s|<1}}_{|l|>1}
\frac{\sin^2 \frac{ \Delta E_{q,s}^{k,l}t}{2} }
{ (\Delta E_{q,s}^{k,l}/2) ^2~ t}\delta_{k+l,q+s},
\ee
\be
\tilde{\Gamma}_k(t)\equiv\sum_p \tilde{\Gamma}_{k \to p}(t)
\ee
and $\Delta E_{q,s}^{k,l}$ is a difference between final and initial kinetic energy of a pair of colliding particles (the impurity and a host fermion):
\be
\Delta E_{q,s}^{k,l} \equiv \frac{k^2-q^2}{2\mi}+\frac{l^2-s^2}{2 \mh}.
\ee

For times considerably greater than the Fermi time $t_{\rm F}$ one can replace
\be
\frac{\sin^2 \frac{ \Delta E_{q,s}^{k,l}t}{2} }
{ (\Delta E_{q,s}^{k,l}/2) ^2~ t}\to 2\pi \delta(\Delta E_{q,s}^{k,l})
\ee
which is essentially the treatment leading to the Fermi's golden rule. One the other hand Eq.~\eqref{naive} suggests that $w_k(t)$ varies slowly on time scales
$\sim t_{\rm F}$, hence the l.h.s of Eq. \eqref{naive} can be replaced by the time derivative at $t \simeq 0$.  Assuming further that the evolution is Markovian, one extends the equation to $t>0$.  This way
one obtains the Boltzmann equation:
\be\label{Boltzmann equation}
\frac{\partial}{\partial t} w_k(t)=-\Gamma_k w_k(t)+\sum_{q} \Gamma_{q \rightarrow k} w_{q}(t),
\ee
where
\be\label{partial width}
\Gamma_{q\to k} =
\frac{\gamma^2}{\pi^2  m_h L} \frac{\theta_{\Omega}\bigl(q,k\bigr)}{|q-k|}\,,
\ee
is the partial width and $\Gamma_k \equiv \sum_{p} \Gamma_{k \rightarrow p} $ is the total width, where $\theta_{\Omega}\bigl(q,k\bigr)$ means that point $\bigl(q,k\bigr)$ lays in domain $\Omega$. The explicit expressions for the dimensionless total width
\be\label{total dimensionless width}
{\bf \Gamma}_k \equiv \left( \frac{\gamma^2 \kF^2}{ 2\pi^3 \mh}\right)^{-1}\Gamma_k=
\left|\log\frac{k-d(k)}{k-u(k)}\right| \theta(|k|-q_0),
\ee
calculated in the thermodynamic limit, are presented in Table \ref{tab width} (note that we have restored $\kF$ in this formula).

Rigorous diagrammatic derivation of the Boltzmann equation \eqref{Boltzmann equation} provides its applicability conditions which read \cite{Burovski2013,Gamayun2014keldysh}
\be\label{applicability conditions}
\gamma^2 \ll 1,~~~~~|1-\eta|\gg\gamma.
\ee
While the former condition is merely an obvious weak-coupling requirement, the latter one does not follow from the above heuristic treatment and deserves a special discussion. This condition implies that the the Boltzmann equation \eqref{Boltzmann equation} is invalid when the masses of the impurity and the host are equal. As was already pointed in the previous section, the kinematics is very special in the equal masses case. Indeed, a pairwise scattering leads merely to the exchange of momenta between an impurity and a host particle. Therefore after the {\it first} scattering the impurity drops below $\kF$ and acquires the same velocity as the created hole. Further scattering processes are not forbidden by the conservation laws and Pauli principle, in contrast to the case of nonequal masses. As a consequence the diagrammatic expansion appears to be dominated by ladder diagrams which modify the form of the evolution equation. The second applicability condition ensures that such diagrams can be disregarded.


\subsection{Iterative solution}
Thanks to a peculiar kinematics discussed in Sec. \ref{sec kinematics}, the Boltzmann equation can be solved by iterations. Let us concentrate on the  initial condition \eqref{initial condition}.  First, obviously,
\be
w_{p_0\to k}(t)=\delta_{p_0 k} \qquad {\rm for} \qquad |p_0|\leq q_0.
\ee
Further,
if $|p_0|$  lies in the interval $(q_{n-1},q_n]$ (and thus is less than $q_\infty$), then the Boltzmann equation can be solved {\it exactly} in $(n+1)$ iterations. Moreover, lower number of iterations, $l \leq n,$ provide an exact solution for $w_{p_0\to k}(t)$ with $|k|>q_{n-l}$. We first exemplify this statement by providing the exact solutions for $n=1,2$ and then outline a general iterative procedure.

The solution of Eq. \eqref{Boltzmann equation} for initial conditions \eqref{initial condition} with $q_0 < |p_0| \leq q_1$ reads
\be\label{Boltzmann solution 1}
w_{p_0\to k}(t) =
\left\{
\begin{array}{ll}
 e^{-\Gamma_{p_0} t} \delta_{p_0 k}, & |k| > q_0,\\
\frac{\Gamma_{p_0 \rightarrow k}}{\Gamma_{p_0}}(1-e^{-\Gamma_{p_0} t}), & |k|\leq q_0.
\end{array}
\right.
\ee
The solution of Eq. \eqref{Boltzmann equation} for initial conditions \eqref{initial condition} with  $q_1 < |p_0| \leq q_2$ reads
\be\label{Boltzmann solution 2 heavy}
\begin{array}{l}
w_{p_0\to k}(t) =\\
\left\{
\begin{array}{ll}
 e^{-\Gamma_{p_0} t} \delta_{p_0 k}, & |k| > q_1, \\
\frac{\Gamma_{p_0 \rightarrow k}}{\Gamma_{p_0}-\Gamma_k}(e^{-\Gamma_k t}-e^{-\Gamma_{p_0} t}), & q_0 < |k| \leq q_1, \\
\frac{\Gamma_{p_0 \rightarrow k}}{\Gamma_{p_0}}(1-e^{-\Gamma_{p_0} t})+ \\
\sum\limits_{q_0 < |q| \leq q_1} \frac{\Gamma_{p \rightarrow q}\Gamma_{q \rightarrow k}}{\Gamma_{p_0}-\Gamma_{q}} \left(
\frac{1-e^{-\Gamma_{q} t}}{\Gamma_{q}}-\frac{1-e^{-\Gamma_{p_0} t}}{\Gamma_{p_0}}
\right), &
 |k|<q_0 \\
\end{array}
\right.
\end{array}
\ee
in the heavy impurity caseand
\be\label{Boltzmann solution 2 light}
\begin{array}{l}
w_{p_0\to k}(t) =
\\
\left\{
\begin{array}{ll}
 e^{-\Gamma_{p_0} t} \delta_{p_0 k}, & |k| > q_1,
\\
\Gamma_{p_0 \rightarrow k} \, t \, e^{-\Gamma_{p_0} t}, & q_0 < |k| \leq q_1,
\\
\frac{\Gamma_{p_0 \rightarrow k}}{\Gamma_{p_0}}(1-e^{-\Gamma_{p_0} t})+
\\
\sum\limits_{q_0 < |q| \leq q_1} \frac{\Gamma_{p \rightarrow q}\Gamma_{q \rightarrow k}}{\Gamma_{p_0}^2}
\left(
1-e^{-\Gamma_{p_0} t}(1+ \Gamma_{p_0} t)
\right), &
 |k|<q_0
\end{array}
\right.
\end{array}
\ee
in the light impurity case.
The difference between former and latter comes from the fact that $\Gamma_k$ is constant for $k\in[q_1,q_{\infty}]$ in the case of a light impurity; formally Eq. \eqref{Boltzmann solution 2 light} can be obtained from  Eq. \eqref{Boltzmann solution 2 heavy} by taking the limit $\Gamma_k,\Gamma_q\rightarrow \Gamma_{p_0}$.

The above sequence of solutions can be systematically  continued to account for an arbitrary $p_0$.
To this end we introduce a new function $\w_{p_0\to k}$ such that
\be
w_{p_0\to k}(t) = e^{-\Gamma_k t} \w_{p_0\to k}(t)
\ee
and rewrite  the Boltzmann equation (\ref{Boltzmann equation})
in terms of this new function:
\be\label{Boltzmann equation rewritten}
\frac{\partial}{\partial t} \w_{p_0\to k}(t)=\sum_q   \w_{p_0\to q}(t)\,\Gamma_{q \rightarrow k}\, e^{-(\Gamma_{q}-\Gamma_{k}) t},
\ee
or
\be\label{Boltzmann equation rewritten integral}
 \w_{p_0\to k}(t)=\delta_{p_0 k}+\sum_q  \int_0^t dt' \w_{p_0\to q}(t')\, \Gamma_{q \rightarrow k} \, e^{-(\Gamma_{q}-\Gamma_{k}) t'} .
\ee
The above equation can be solved iteratively thanks to the fact that only terms with $|q|>|k|$ contribute to the sum on the r.h.s. (according to the expression \eqref{partial width} for the partial width $\Gamma_{q \rightarrow k}$), and therefore one can calculate $\w_{p_0\to k}$  consequentially, starting from the largest $k=p_0$. Indeed, consider $p_0\in (q_{n-1},q_n]$. The first step is to note that $\w_{p_0\to k}(t)=\delta_{p_0 k}$ for $k\in (q_{n-1},q_n]$. Assume now that after $l-1$ steps we know $\w_{p_0\to k}(t)$ for $|k|>q_{n-l+1}$. We substitute it to the r.h.s. of Eq. \eqref{Boltzmann equation rewritten integral} to obtain $\w_{p_0\to k}(t)$ for $|k|\in (q_{n-l},q_{n-l+1}]$, which constitutes the $l$'th step. Evidently, the procedure ends in $(n+1)$ steps. Solutions \eqref{Boltzmann solution 1}, \eqref{Boltzmann solution 2 heavy} and \eqref{Boltzmann solution 2 light} are obtained through this procedure.

Note that the above procedure can be somewhat improved. Indeed, at the first step one can substitute the interval $(q_{n-1},q_n]$ by the interval $(\max\{|u(p_0)|,|d(p_0)|\},p_0]$, and proceed analogously at further steps.

In the case $p_0>q_\infty$ an exact solution can not be obtained in finite number of steps, however Eq. \eqref{Boltzmann equation rewritten integral} can be used to obtain an approximate solution.

Finally, let us write down the solution to the Boltzmann equation (\ref{Boltzmann equation}) for a more general initial condition: $w_k(0)=0$ for $|k|>q_1.$ The solution reads
\be
\begin{array}{l}
w_k(t)= \\
\left\{
\begin{array}{ll}
 e^{-\Gamma_k t} w_k(0), & |k| > q_0,\\
w_k(0)+ \sum\limits_{q_0<|q|\leq q_1} (1-e^{-\Gamma_{q} t})\frac{\Gamma_{q \rightarrow k}}{\Gamma_{q}} w_{q}(0), & |k|\leq q_0.
\end{array}
\right.
\end{array}
\ee
The simplicity of the solution is due to the fact that for a fixed $k$  either the first or the second term on the r.h.s. of Boltzmann equation (\ref{Boltzmann equation}) vanishes.

\section{\label{sec asymptotic distribution}Equilibrium state in the absence of force}

\begin{figure*}[t]
\includegraphics[width=\linewidth]{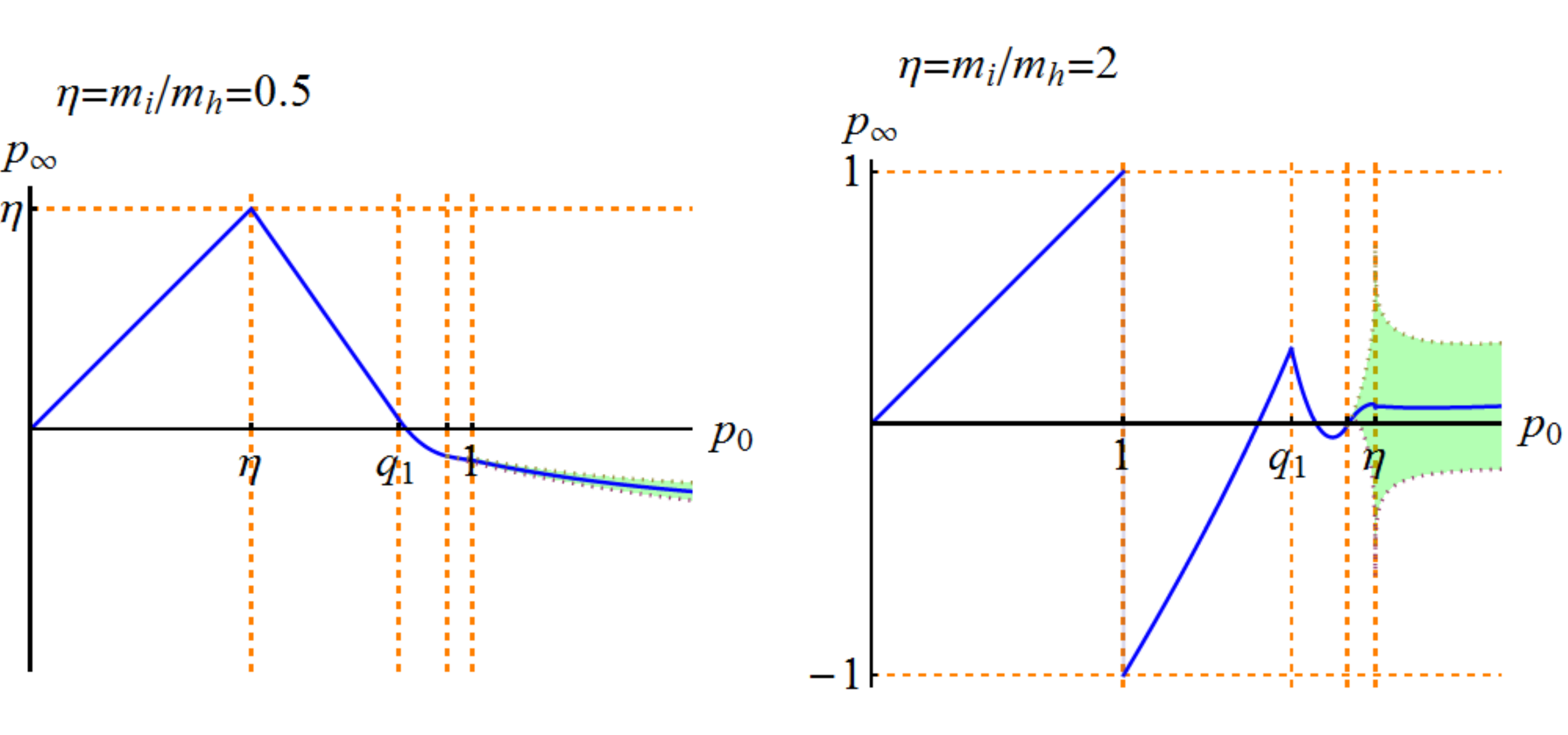}
\caption{\label{fig momentum nonequal masses} The asymptotic impurity momentum $p_\infty$ as a function of the initial  impurity momentum $p_0$ when impurity is lighter (left) and heavier (right) than a host particle. Solid blue line -- an  iterative solution (two iterations) of Eq. (\ref{integral eq for asymptotic momentum}). This solution is exact below $q_2$ (not shown) and approximate above this point.  Shaded area (green online) represents the maximal error: the exact solution of Eq. (\ref{integral eq for asymptotic momentum}) lies inside this area. Notice much better convergence of iterations in the light impurity case. }
\end{figure*}

As is shown in appendix \ref{appendix proof}, Eq. (\ref{Boltzmann equation})
leads to the following integral equation on asymptotic distribution:
\begin{align}\label{integral eq for asymptotic distr}
\begin{array}{l}
w_{p_0\rightarrow k}^\infty = \\
\left\{
\begin{array}{ll}
\theta(q_0-|k|)\left( {\cal P}^{(1)}_{ p_0\rightarrow k}+\sum\limits_{q\in {\cal R}(p_0)} {\cal P}^{(1)}_{ p_0\rightarrow q} w_{q\rightarrow k}^\infty\right), & |p_0|>q_0,\\
\delta_{p_0 k} , & |p_0|<q_0,
\end{array}
\right.
\end{array}
\end{align}
where
\be\label{Prob1}
{\cal P}^{(1)}_{ p_0\rightarrow k} = \Gamma_{p_0\to k}/\Gamma_{p_0}
\ee
is the probability that the impurity changes its momentum from $p_0$ to $k$ in a {\it single} scattering event, and ${\cal R}(p_0)\equiv [d(p_0),u(p_0)]\setminus[-q_0,q_0]$ is a kinematically determined integration region.


Eq. (\ref{integral eq for asymptotic distr}) has a clear semiclassical meaning. In the case $|p_0|<q_0$ any scattering is kinematically forbidden, as discussed in Sec. \ref{sec kinematics}, and the relaxation does not occur. Consider the case $|p_0|>q_0$. After the first scattering a momentum of the impurity
can be either below or above~$q_0.$ In the former case the evolution stops, which is represented by the first term in the large brackets in Eq. \eqref{integral eq for asymptotic distr}; in the latter case the scattering process iterates,  which is represented by the second term in the large brackets in Eq. \eqref{integral eq for asymptotic distr}.

One can easily write down the solution of Eq. \eqref{integral eq for asymptotic distr}. Consider $|p_0|\in (q_{n-1},q_n]$, $n \geq 1$. Then the solution reads
\be\label{solution for asymptotic distr}
w_{p_0\rightarrow k}^\infty = \theta(q_0-|k|) \sum_{j=1}^n {\cal P}^{(j)}_{ p_0\rightarrow k},
\ee
where
\be\label{Prob(j)}
{\cal P}^{(j)}_{ p_0\rightarrow k}=\sum_{|q|>q_0}{\cal P}^{(1)}_{ p_0\rightarrow q} {\cal P}^{(j-1)}_{ q\rightarrow k}
\ee
is the probability of the evolution path which leads from $p_0$ to $k$ in exactly $j$ scattering events.
  In case of $|p_0|>q_\infty$ the upper limit of summation, $n$, should be substituted by $\infty$.

Calculating the first moment of the distribution~\eqref{integral eq for asymptotic distr} (with respect to $k$) we find an
integral equation for the asymptotic
momentum,
\be\label{integral eq for asymptotic momentum}
p_\infty (p_0) =
\left\{
\begin{array}{ll}
p_\infty^{(1)} (p_0)
+\sum\limits_{q \in{\cal R}(p_0)}  {\cal P}^{(1)}_{ p_0\rightarrow q} p_\infty
(q), & |p_0|>q_0,\\
p_0 , & |p_0|<q_0,
\end{array}
\right.
\ee
where
$$p_\infty^{(1)} (p_0)\equiv \sum_{k=-q_0}^{q_0} k ~{\cal P}^{(1)}_{
p_0\rightarrow k}.
$$
The solution of this equation is obtained from the solution \eqref{solution for asymptotic distr} of the parent equation~\eqref{integral eq for asymptotic distr}. If $|p_0|\in (q_{n-1},q_n]$, $n \geq 1$,  then
\be\label{solution for asymptotic momentum}
p_\infty (p_0) =  \sum_{j=1}^n p_\infty^{(j)} (p_0),
\ee
where
\be\label{p(j)}
p_\infty^{(j)} (p_0) \equiv \sum_{|k| \leq q_0} k~ {\cal P}^{(j)}_{ p_0\rightarrow k}=\sum_{|q|>q_0}{\cal P}^{(1)}_{ p_0\rightarrow q} p_\infty^{(j-1)} (q).
\ee
In the case of $|p_0|>q_\infty$ the upper limit of summation, $n$, should be substituted by $\infty$.

Observe that $\gamma^2$ cancels out from  the probability  ${\cal P}^{(j)}_{ p_0\rightarrow k}$. As a result, the asymptotic distribution $w_{p_0\rightarrow k}^\infty$ given by Eq. \eqref{solution for asymptotic distr} and the infinite-time momentum $p_\infty$ given by Eq. \eqref{solution for asymptotic momentum} do not depend on the coupling constant at all. This remarkable fact will be discussed in what follows.

The recurrent nature of expressions for  ${\cal P}^{(j)}_{ p_0\rightarrow k}$ and $p_\infty^{(j)} (p_0)$, Eqs. \eqref{Prob(j)} and \eqref{p(j)} correspondingly, implies that in fact the solutions \eqref{solution for asymptotic distr} and \eqref{solution for asymptotic momentum} are constructed iteratively, in resemblance to the solution of the Boltzmann equation \eqref{Boltzmann equation}. Namely, the calculation of the $j$'th terms on the r.h.s. of Eqs.  \eqref{solution for asymptotic distr} and \eqref{solution for asymptotic momentum} constitutes the $j$'th  step of iteration.

Although formally we have solved the integral equations \eqref{integral eq for asymptotic distr} and \eqref{integral eq for asymptotic momentum}, practical implementation of the full solutions \eqref{solution for asymptotic distr} and \eqref{solution for asymptotic momentum} can be tedious because of a large or even infinite number of terms in the corresponding sums. A natural question arises whether one can truncate these sums at some small $j$ to obtain reliable approximation to the corresponding exact solutions. The answer appears to be affirmative. Below we demonstrate this for the infinite-time momentum.

We define an approximate solution
\be
p_\infty^{(1,...,l)} (p_0) \equiv \sum_{j=1}^l p_\infty^{(j)} (p_0)
\ee
with $l<n$ and study the difference between it and the exact solution \eqref{solution for asymptotic momentum}. This difference appears to be  bounded from above according to
\be\label{bound}
|p_\infty (p_0)-p_\infty^{(1 \dots l)} (p_0)| \leq \left(1-\sum_{j=1}^l {\cal P}^{(j)}_{p_0}\right) q_0,
\ee
where
\be\label{Prob(j) total}
{\cal P}^{(j)}_{p_0}\equiv \sum_{k=-q_0}^{q_0} {\cal P}^{(j)}_{
p_0\rightarrow k}=
\sum_{|q|>q_0}{\cal P}^{(1)}_{ p_0\rightarrow q} {\cal P}^{(j-1)}_{ q}
\ee
is the probability that the impurity experiences {\it exactly} $j$ scatterings before its momentum drops below $q_0$. This bound again has a clear semiclassical meaning: the maximal error in momentum after $l$ iterations does not exceed the maximal possible infinite-time momentum, $q_0$, times the probability that the impurity experiences {\it more} than $l$ scatterings before its momentum drops below $q_0$.

According to the estimate \eqref{bound}, the upper bound of the discrepancy $|p_\infty (p_0)-p_\infty^{(1 \dots l)} (p_0)|$ monotonically approaches zero with increasing $l$ for any fixed $p_0$, except the case $\eta>1,~p_0=\eta$. In the latter case, which is discussed in more details in what follows, the total width $\Gamma_{p_0}$ diverges and thus the probabilities ${\cal P}^{(j)}_{p_0}$ are not well defined. Apart from the above-mentioned special case, the estimate \eqref{bound} allows to explicitly control the convergence of the iterative procedure which leads to the solution \eqref{solution for asymptotic distr} for the infinite-time momentum $p_\infty(p_0)$.

Let us describe in more detail the first iteration. The probability ${\cal P}^{(1)}_{p_0}$ reads
\begin{align}\label{Prob1 explicit}
\begin{array}{l}
{\cal P}^{(1)}_{p_0}=\\
\left\{
\begin{array}{ll}
0, & |p_0|\leq q_0,\\
{\bf \Gamma}_{p_0}^{-1}
\left|\log\left|\frac{\Delta_d(p_0)}{\Delta_u(p_0)}\right|\right|, & |p_0|\in \left(q_0,\frac{3\eta+q_0^2}{|\eta-1|}\right],\\
0, & |p_0|> \frac{3\eta+q_0^2}{|\eta-1|},\\
\end{array}
\right.
\end{array}
\end{align}
where $\Delta_d(p)=p-\max\{d(p),-q_0\}$ and $\Delta_u(p)=p-\min\{u(p),q_0\}$ are the maximum and minimum momentum transferred in scattering event, correspondingly.
In particular, in the case $|p_0|>\frac{3\eta+q_0^2}{|\eta-1|}$   the momentum of the impurity is kinematically prohibited to drop below $q_0$ in a single scattering. Note that for $q_0<p_0\leq q_1$ Eq.~\eqref{Prob1 explicit} amounts to ${\cal P}^{(1)}_{p_0}=1$.
We recall that the expression for ${\bf \Gamma}_{p_0}$ is given by Eq.~\eqref{total dimensionless width} and can be read from Table \ref{tab width}.
The contribution of the first iteration to the asymptotic momentum reads
\begin{align}\label{p1}
\begin{array}{l}
p_\infty^{(1)} (p_0)=\\
\left\{
\begin{array}{lll}
0, & |p_0|\leq q_0, & \\
\multicolumn{2}{l}{p_0 {\cal P}^{(1)}_{p_0}-
(\Delta_{d}(p_0)-\Delta_u(p_0))/{\bf \Gamma}_{p_0},} & \\
\multicolumn{3}{r}{|p_0|\in \left(q_0,\frac{3\eta+q_0^2}{|\eta-1|}\right],}\\
0, & |p_0|> \frac{3\eta+q_0^2}{|\eta-1|}. &\\
\end{array}
\right.
\end{array}
\end{align}


The infinite-time momentum of the impurity  as a function of its initial momentum, calculated in two iterations, is displayed in Fig. \ref{fig momentum nonequal masses} The plotted curve represents the exact solution of Eq.~\eqref{integral eq for asymptotic momentum} for $|p_0| \leq q_2$ and an approximate solution for larger values of $p_0$. In the latter case maximal error, calculated according to Eq.~\eqref{bound}, is also shown.

The solution of Eq.~(\ref{integral eq for asymptotic momentum}) as a function of $p_0$ has non-analyticities
at $p_0=q_0,q_1,\dots,q_\infty.$ The most prominent discontinuity occurs for $\eta>1$ at $|p_0|=q_0=1$. In this case, according to Eqs.~\eqref{solution for asymptotic momentum} and~\eqref{p1},
\begin{align}\label{asymptotic momentum heavy impurity}
\begin{array}{l}
p_{\infty}(p_0)=\\
\left\{
\begin{array}{ll}
p_0, & |p_0|<1 \\
-1 + \frac{\eta^2 + 1}{\eta^2-1 }(p_0 - 1) + O\left((p_0-1)^2\right),& p_0>1,
\end{array}
\right.
\end{array}
\end{align}
One can see that the infinite-time momentum as a function of $p_0$ flips sign and $p_0=1$. The reason is that backscattering with momentum transfer $2k_{\rm F}$ is the only process allowed kinematically at this point, see Fig. \ref{fig kinematics}.

Another type of singularity shows up at $p_0=\eta$. This is the only point at which the forward scattering (scattering with zero momentum transfer) is kinematically allowed. In the heavy impurity case, $\eta>1$, the forward scattering contribution leads to the divergence of the total width in the thermodynamic limit, see Table \ref{tab width}, and precludes the convergence of the series on the r.h.s. of Eqs.~\eqref{solution for asymptotic distr} and ~\eqref{solution for asymptotic momentum}. The convergence of the series in the vicinity of this point is non-uniform in $p_0$. The physical picture behind this singularity is as follows: In the thermodynamic limit, the impurity scatters infinitely many times with infinitely small momentum transfer before its momentum drops below $q_0$.

We believe that the non-analyticities at $p_0=q_0,q_1,\dots,q_\infty$ are the artifacts of the weak coupling approximation which underlies the Boltzmann equation~\eqref{Boltzmann equation}. Consider the {\it exact} asymptotic momentum as a function of the initial momentum and coupling constant and imagine that it is expanded in a series with respect to $\gamma$. Eqs. \eqref{integral eq for asymptotic momentum} are nothing else than the equations for the leading  (i.e. $O(1)$) term of this expansion. We expect that taking into account higher order terms would smoothen the above-mentioned non-analyticities.

In general, our approach based on the semiclassical kinetic theory is not suitable for obtaining next-to-leading in $\gamma$ terms. However, in the case when $|p_0|<q_0$ and, moreover,  $p_0$ is sufficiently far from $q_0$, one can drive higher order corrections from the straightforward perturbation theory discussed in Sec. \ref{subsec derivation of Boltzmann equation}. In particular, the asymptotic momentum reads
\be\label{pinfty with quantum corrections}
p_\infty=p_0 -\frac{\gamma^2 \eta^2 }{\pi^4 }  \frac{\log\frac{1+p_0}{1-p_0}}{\eta^2 -p_0^2}.
\ee
This expression is valid
if $q_0-p_0\gg\Delta p $ where
\be\label{dp}
\begin{array}{ll}
\Delta p =\exp\left( - \frac{\pi^4 } {\gamma^2 } \frac{\eta^2 -1}{\eta^2 }\right), & \quad\eta>1\\
\Delta p =  \frac{\gamma^2 \eta }{\pi^4 }, & \quad \eta<1
\end{array}\,,
\ee
which ensures that correction on the r.h.s. of \eqref{pinfty with quantum corrections} is small.

\section{\label{sec force}Impurity under the action of a constant force}

\subsection{\label{sec evolution with force}Time evolution in the presence of a force}

\begin{figure*}[t]
\includegraphics[width=  \linewidth]{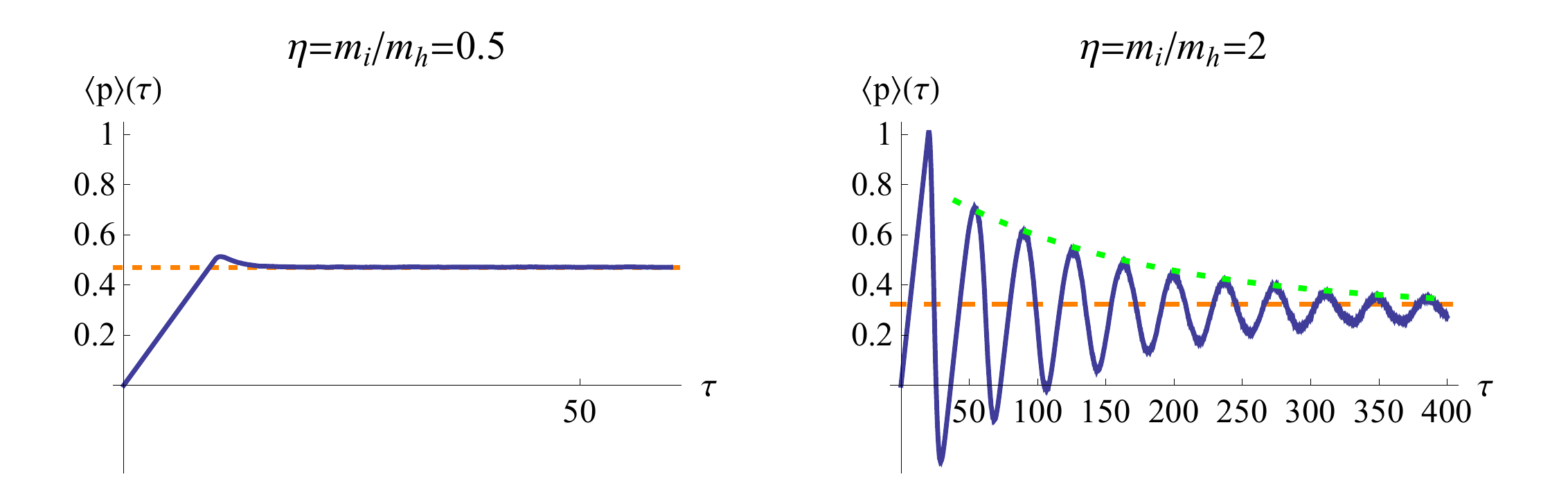}
\caption{\label{fig force}
Time evolution of the average momentum of the impurity moving under the action of a constant force. Initially the impurity is at rest. Solid curves -- numerical solutions of the Boltzmann equation \eqref{Force2} with $f=0.05$.
Dashed lines -- asymptotic momentum according to Eqs. (\ref{averageP2}) (light impurity) and (\ref{averageP1}) (heavy impurity). Oscillations are present in the heavy impurity case and absent in the light impurity case. The attenuation of the amplitude of the oscillations  is shown by dotted line, see Eq. \eqref{decrement}.
}
\end{figure*}

The Boltzmann equation in the presence of a constant force $F$ reads
\be\label{Force1}
\frac{\partial w_k(t)}{\partial t} + F \frac{\partial w_k(t)}{\partial k}=-\Gamma_k w_k(t)+\sum_{q} \Gamma_{q \rightarrow k} w_{q}(t).
\ee
Without loss of generality we assume $F>0$. We chose to treat $k$ as a continuous variable in the present section and replace all sums by integrals according to $\sum_q \rightarrow L/2\pi \int dq$.\footnote{This allows us to introduce the derivative $\partial/\partial k$ and to treat the evolution operator $\hat L$ in Eq. \eqref{Force2} as an integro-differential operator.} We replace probability $w_k$ by probability distribution $w(k)= w_k L/(2\pi)$. Also it is useful to introduce   re-scaled dimensionless force and time:
\be\label{dimensionless force and time}
f \equiv 2\pi^3  m_h F/(\gamma^2\kF^3),~~~\tau \equiv t/\left( \frac{2\pi^3 \mh}{\gamma^2\kF^2} \right),
\ee
where we have restored $\kF$. Note that the coupling constant $\gamma$ enters the scaling factors.
Finally, one can rewrite Eq. \eqref{Force1} in the following form:

\be\label{Force2}
\frac{\partial }{\partial \tau}w(k,\tau)=\hat L\, w(k,\tau),
\ee
where the operator $\hat L$ is defined according to
\begin{align}\label{operator}
&\hat L\, w(k,\tau)=-f \frac{\partial w(k,\tau)}{\partial k} \nonumber\\
& - w(k,\tau) \int dq \frac{\theta_{\Omega}(k,q)}{|k-q|}+\int dq \frac{\theta_{\Omega}(q,k)}{|k-q|}w(q,\tau).
\end{align}

In addition to the applicability conditions discussed in Sec. \ref{sec Boltzmann equation} one more condition the required for the Boltzmann equation with force \eqref{Force1}. Indeed, one needs to make sure that the time $ \Delta p/ F$ which impurity spends in the $\Delta p$-vicinity of $q_0$  \eqref{dp} is much less than the mean free time. This entails the following applicability conditions
\begin{align}\label{not too small force heavy}
& f \gg e^{ - \frac{\pi^4 } {\gamma^2 }\frac {\eta^2 -1}{\eta^2 }}, & \eta>1,
\\
\label{not too small force light}
& f \gg  \gamma^2\eta, & \eta<1.
\end{align}
We note that in the heavy impurity case the lower bound for $f$ is exponentially suppressed in the limit of small~$\gamma$.

\subsection{\label{sec oscillations}Backscattering oscillations and saturation without oscillations}
Eq. \eqref{Force2} can be easily solved numerically, either directly or by the Monte-Carlo method. In the latter case a trajectory of the impurity subject to a random scattering process is simulated.  The evolution of the momentum of the impurity obtained from the Monte Carlo simulation is shown in Fig. \ref{fig force}. One can see that the motion of the heavy impurity exhibits oscillations. This effect has been predicted in \cite{Gangardt2009,schecter2012dynamics,Gangardt2012} (see, however, discussion below). In contrast, in the light impurity case oscillations are absent.

This behavior can be easily understood from kinematical arguments. Indeed, consider an impurity initially at rest. The force accelerates it freely until its momentum reaches $q_0$. Then the impurity acquires a chance to scatter on a host particle. As is clear from Fig. \ref{fig kinematics}, in the heavy impurity case the scattering process transfers the impurity to an opposite edge of the Fermi sea, after which it starts to freely accelerate again until its momentum reaches $q_0$, and so on. A term ``backscattering oscillations'' has been proposed for this phenomenon in Ref. \cite{Lychkovskiy2014}. Obviously, in the limit of small $f$ the period of the backscattering oscillations reads $\tau_{\rm cycle}=2/f$, or, restoring physical time, force and momentum,
\be
t_{\rm cycle}=2\kF/F.
\ee

In contrast, if the impurity is light then the dominant scattering processes are those with small momentum transfer. They tend to freeze the momentum of the impurity in the vicinity of $q_0$, which leads to a rapid saturation of the momentum, as shown on Fig. \ref{fig force}.

The backscattering oscillations get damped since the distribution gets  broadened by scattering events. In the long run the system approaches a steady state. The steady state and the damping rate can be found by solving the eigenvalue problem for the operator $\hat L$. The zero eigenvalue of $\hat L$  corresponds to the steady state which is analyzed in detail in the next subsection.  Here we briefly comment on the whole spectrum of $\hat L$ focusing on the heavy impurity case. For $f \ll 1$ one can show that the eigenvalues $\lambda$ satisfy the following equation:
\begin{align}\label{equation for eigenvalues}
&f=\int_{-1}^{1} dk \int_{1}^{\infty} dq\,\frac{\theta_{\Omega}(q,k)}{|k-q|}
e^{
-\frac{1}{f} \left(\int\limits_{1}^{q} ds\, {\bf \Gamma}(s)+\lambda(q-p) \right)
}.
\end{align}
All solutions of \eqref{equation for eigenvalues}, except $\lambda=0$, have negative real parts.
The smallest nonzero $|{\rm Re} \lambda|$ determines the damping rate at large times. This way we find that for $\eta>1$ the amplitude of backscattering oscillations $A(\tau)$ is damped as
\be\label{decrement}
A(\tau)\sim\exp\left(- \pi^2 \frac{1+\eta^2/3-\pi/4}{\eta(\eta^2-1)} f^2 \tau\right),~~~f \ll 1\,.
\ee
This formula compares well with the numerical solution of Eq. \eqref{Force2}, see Fig. \ref{fig force}.

Our results on the dynamics of the impurity under the action of a small constant force seem to be in apparent contradiction to the results reported in Refs. \cite{Gangardt2009,schecter2012dynamics,Gangardt2012}. In the latter works the quasi-Bloch oscillations of the momentum of the impurity has been predicted to occur for any mass of the impurity. Thus the disagreement between their and our results is most dramatic in the case of light impurity when we do not find any oscillations.  On the other hand, the oscillations which we find in the case of heavy impurity resemble the quasi-Bloch oscillations of Refs. \cite{Gangardt2009,schecter2012dynamics,Gangardt2012}, although the physics underlying the two types of oscillations seems to be rather different: The quasi-Bloch oscillations of \cite{Gangardt2009,schecter2012dynamics,Gangardt2012} is a coherent effect while we obtain the oscillations from the Boltzmann equation describing incoherent scattering events.

Strictly speaking, the results of the present paper and of Refs. \cite{Gangardt2009,schecter2012dynamics,Gangardt2012} are not necessarily in conflict since the formal range of validity of the former is restricted to not very small forces, see Eqs. \eqref{not too small force heavy}, \eqref{not too small force light}, while the latter are obtained in the limit of vanishing force.  The question whether crossing over these two parametric regimes can eliminate the controversy has triggered an intensive discussion \cite{Lychkovskiy2014,schecter2014comment,Gamayun2014reply} which has emerged after the first version of the present paper had been made available as an electronic preprint. To the moment the question is not not resolved completely and requires further careful analysis.

\subsection{Steady state}

The properties of the steady state of the system (state at infinite time) are of major interest. We are able to study them analytically. The steady state, if it exists, is a solution of the equation
\be\label{equation for steady state}
\hat L\, w(k)=0.
\ee
Here and in the rest of this section we do not study any time dependance and use a notation $w(k)$ for a steady state.

The symmetric form of the r.h.s. of Eq. \eqref{operator} allows us to present the steady state average of any quantity as
\begin{multline}
 \int dp\, w(p) O(p) =
  \\ \label{average}
  =\frac{1}{f}\int dk \int dq \,w(q) \frac{\theta_{\Omega}(q,k)}{|k-q|} \int\limits_{k}^q ds\, O(s).
\end{multline}
This representation is very useful because it shows that any average can be expressed through probabilities outside the domain $[-q_0,q_0]$. To emphasize this we introduce special notations:
\be
w(k) =
\begin{cases}
 \chi^\>(k) &\mbox{if }  k\ge q_0 \\
 \chi^\<(k) & \mbox{if } k\le -q_0.
\end{cases}
\ee

The probability distribution  is determined by the Boltzmann equation up to a multiplicative constant. It should be deduced from the normalization condition, which according to Eq. \eqref{average} reads 
\begin{multline}\label{normalization condition}
1  = \frac{1}{f}\int_{q_0}^{\infty} dq  \, \chi^\>(q) (u_\>(q)-d_\>(q)) \\- \frac{1}{f}\int^{-q_0}_{-\infty} dq  \, \chi^\<(q) (u_\<(q)-d_\<(q)).
\end{multline}
In what follows we calculate unnormalized distribution and divide all averages by the r.h.s. of the above equation.

The average momentum according to Eq. \eqref{average} reads
\begin{align}
\nonumber
& p_\infty(f) =\\
\nonumber
& \frac{1}{4f}\int\limits_{q_0}^{\infty} dq  \, \chi^\>(q) (u_\>(q)-d_\>(q))(u_\>(q)+d_\>(q)+2q) \\
\label{averageP}
-& \frac{1}{4f}\int\limits^{-q_0}_{-\infty} dq  \, \chi^\<(q) (u_\<(q)-d_\<(q))(u_\<(q)+d_\<(q)+2q).
\end{align}
Now the task is to calculate $\chi^\>(s)$ and $\chi^\<(s)$.
The results for $\eta>1$ and $\eta<1$
appear to be quite different.   Below we consider these cases separately.

\subsubsection{Heavy impurity}

\begin{figure*}[t]
\includegraphics[width= 0.4 \linewidth]{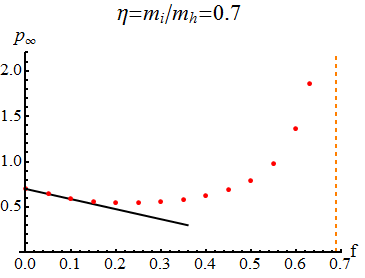}
\includegraphics[width= 0.4 \linewidth]{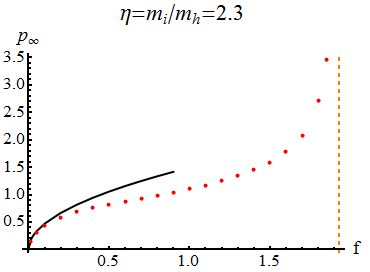}
\caption{\label{fig steady state with force}
Average momentum of a steady state, $p_\infty$, {\it vs} dimensionless force $f$  applied to the impurity (see Eq. \eqref{dimensionless force and time} for definition of $f$).
Black lines --  asymptotical expressions in the small force limit  according to  Eqs. (\ref{averageP2}) and (\ref{averageP1}) for light and heavy impurity, correspondingly.
Red dots -- numerical solution of the steady state Boltzmann equation \eqref{equation for steady state}.
 The steady state momentum diverges at a critical force $f_{c1}$ (marked by a dashed vertical line), which is given by Eqs.  \eqref{critical forces light} and \eqref{critical force} for light and heavy impurity, correspondingly.
}
\end{figure*}

When the impurity is heavier that host particle Eq. \eqref{equation for steady state} leads to
\be\label{chi1}
f\frac{d\chi^\>(k)}{dk} = -\chi^\>(k) {\bf \Gamma}(k) + \int\limits^{d^{-1}_\>(k)}_{u^{-1}_\>(k)} ds \frac{\chi^\>(s)}{s-k},
\ee
\be\label{hatchi1}
f\frac{d\chi^\<(k)}{dk}  = -\chi^\<(k) {\bf \Gamma}(k) + \int\limits^{d^{-1}_\<(k)}_{u^{-1}_\<(k)} ds \frac{\chi^\<(s)}{s-k},
\ee
where ${\bf \Gamma}(k)$ equals to ${\bf \Gamma}_k$ (see Eq. \eqref{total dimensionless width}) calculated in the thermodynamic limit and can be read from Table \ref{tab width}.
We see that this is a set of decoupled equations. Moreover, $\chi^\<(k)$ is identically zero (for a positive force), which can be proven by integrating both sides of Eq. \eqref{hatchi1} over $k$ and taking into account non-negativity of the probability distribution.

The existence of the normalizable solution of \eqref{chi1} can be tested by considering large momentum asymptote. Taking into account
\be
{\bf \Gamma}(k) \sim \frac{2\eta}{k},\,\,\,\,\, k\gg 1,
\ee
one looks for an asymptote in the form $\chi^\>(k) \sim k^{-\alpha}$ which gives the following relationship between $f$ and $\alpha$ :
\be\label{f and alpha}
f = \frac{2\eta}{\alpha}\left(1-\left(\frac{\eta-1}{\eta+1}\right)^{\alpha}\right)
\ee
Thus the normalizable distributions exists up to a critical force
\be
f_{c0}\equiv \frac{4\eta}{1+\eta},
\ee
while a finite average momentum --  up to a critical force
\be\label{critical force}
f_{c1} \equiv 4\eta^2/(1+\eta)^2.
\ee

In the limit $f\to 0$ Eq. \eqref{f and alpha} implies $\alpha\to\infty$ and all the weight of $\chi^\>(k)$ is concentrated near $k=1$.
Therefore the integral term in \eqref{chi1} can be neglected in this limit and one obtains
\begin{multline}\label{approximate chi heavy}
\chi^\>(k) \sim \exp\left(
-\frac{1}{f}\int\limits_{1}^{k} ds\, {\bf \Gamma}(s)
\right)= \\
=  \exp\left(
-\frac{1}{f}\left( k\log\frac{(\eta+k)(\eta-1)}{(\eta-k)(\eta+1)}+\eta \log\frac{\eta^2-k^2}{\eta^2-1} \right)\right) \\
\simeq \exp\left(-\frac{\eta(k-1)^2}{f(\eta^2-1)}\right),
\end{multline}
up to a normalization factor. Calculating the normalization factor from \eqref{normalization condition} and substituting
obtained $\chi^\>(k)$ to \eqref{averageP} we find that the average momentum of the steady state for a small force reads
\be\label{averageP1}
p_\infty(f)
 =\frac{1}{2}\sqrt{\frac{\eta^3}{\eta^2-1} \pi f}.
\ee
This approximation is valid
for a force sufficiently  weak to satisfy $\chi^\> (1)\gg \chi^\>(u^{-1}_\>(1))$. Note that $u^{-1}_\>(1)$ is nothing but $q_1$ (see Table \ref{table qn}), therefore this approximation implies that all the probability spreads no far then the "single scattering"~ region. The explicit applicability condition reads
\begin{multline}\label{smallness1}
e^{-\frac{1}{f}\left(2\eta \log\left[\frac{\eta^2+4 \eta-1}{(1+\eta)^2}\right]-\frac{(1-\eta)^2}{1+\eta}\log\left[\frac{\eta^2+4 \eta-1}{\eta^2-1}\right]\right)}\ll 1.
\end{multline}
In limiting cases this amounts to
\be
f\lesssim
\begin{cases}
\eta-1, & \mbox{if }~ \eta-1 \ll 1,\\
4/\eta, & \mbox{if }~ \eta \gg 1.
\end{cases}
\ee
One can see that the approximation \eqref{averageP1} fails both in the limit of equal masses and in the limit of infinitely heavy impurity. Nevertheless it works well for intermediate values of $\eta$ in which case the condition \eqref{smallness1} amounts to $f \lesssim 1/4$.

It is interesting to compare our result \eqref{averageP1} with a prediction of Ref. \cite{schecter2012dynamics}  that the steady state momentum is a linear function of force at small forces.
Although the two results are in apparent contradiction it is possible that square root dependence crosses over to a linear behavior at exponentially small forces determined by the condition \eqref{not too small force heavy}.

\subsubsection{Light impurity}

For $\eta<1$ one obtains from Eq. \eqref{equation for steady state}
\be
f\frac{d\chi^\>(k)}{dk} = -\chi^\>(k) \gamma^>(k)  + \int\limits_{d^{-1}_\<(k)}^{u^{-1}_\<(k)} ds \frac{\chi^\<(s)}{r-s},
\ee
\be
f\frac{d\chi^\<(k)}{dk} = -\chi^\<(k) \gamma^<(k) + \int\limits_{d^{-1}_\>(k)}^{u^{-1}_\>(k)} ds \frac{\chi^\>(s)}{s-k}.
\ee
One can see that is the light impurity case the equations do not decouple.
Repeating large momentum analysis we find following asymptotes:
\be
\chi^>(k)\sim \frac{1}{k^{\alpha}},\,\,\,\chi^<(k)\sim \frac{2\eta}{2\eta+f\alpha}\left(\frac{1-\eta}{1+\eta}\right)^{\alpha}\frac{1}{k^{\alpha}}
\ee
and corresponding relationship between $\alpha$ and $f$:
\be\label{fa}
f = \frac{2\eta}{\alpha}\left(1-\frac{2\eta}{2\eta+f\alpha}\left(\frac{1-\eta}{1+\eta}\right)^{2\alpha}\right).
\ee
This relationship gives the critical values of force below which the static distribution ($f_{c0}$) and finite average momentum ($f_{c1}$) exist:
\be\label{critical forces light}
f_{c0} = \frac{4\eta^{3/2}}{1+\eta},\,\,\,\,\,\,f_{c1} = \frac{2\eta\sqrt{2\eta(1+\eta^2)}}{(1+\eta)^2}
\ee
From equation \eqref{fa} we see that at a small force $\chi^\>(k)\gg \chi^\<(k)$, therefore we again can neglect the $\chi^\<(k)$ contribution. A calculation analogous to one in the heavy impurity case leads to
\be\label{chi light approximate}
\chi^\>(k) \sim \exp\left(-\frac{k}{f}\log\frac{1+\eta}{1-\eta}\right)
\ee
and
\begin{multline}
\label{averageP2}
p_\infty(f)
=
\eta-f \frac{2 \eta^2 }{1-\eta^2}\left(\log\frac{1+\eta}{1-\eta}\right)^{-1},
\end{multline}
where an additional simplification comes from the fact that ${\bf \Gamma}(k)={\rm const}$ for $\eta < k< 1$, see table \ref{tab width}.
The applicability condition for Eqs. \eqref{chi light approximate}, \eqref{averageP2} reads $\chi^\>(q_1)\ll \chi^\>(q_0)$, or, explicitly,
\be\label{smallness2}
\exp\left(-\frac{2\eta}{f}\frac{1-\eta}{1+\eta}\log\frac{1+\eta}{1-\eta}\right)\ll 1.
\ee
It is clear from this condition that the approximation does not work in limiting cases of equal masses or infinitely light impurity, however, it works well for intermediate values of $\eta$, at which case condition \eqref{smallness2} amounts to $f\lesssim 1/4$, similar to the heavy impurity case.

We note the following remarkable qualitative feature of the steady state momentum \eqref{averageP2}.
The function $p_{\infty}(f)$ has a non-vanishing value in the limit $f\to 0$.
Even though this limit is formally inaccessible in our theory  (see Eq. \eqref{not too small force light}), we believe that the jump discontinuity of the steady state momentum as a function of force is physical and is not an artefact of our approximation.
The reason for this is the absence of physical process with large momentum transfer at impurity momentum equal to $q_0$.
This results disagrees with predictions made in \cite{schecter2012dynamics} that the steady state momentum is linear in force at small forces.

\section{\label{sec summary and outlook}Summary and outlook}
To summarize, we apply the Boltzmann kinetic theory to the dynamics of a mobile impurity in a Tonks-Girardeau gas at zero temperature.
We develop the iterative procedure solving the Boltzmann equation. This procedure allows us to obtain explicit analytical results for the impurity momentum distribution in a number of special cases. Moreover, in the general case it provides a controlled approximate method for the calculation of the impurity's average momentum.
Our solution reveals a striking peculiarity of the impurity dynamics. Namely, the system retains the memory about its initial state beyond that imposed by conservation laws, which signals the breakdown of thermalization (see Fig. \ref{fig momentum nonequal masses}). The microscopic reason for this is the kinematical constrains on two-particle scattering processes combined with the dome structure of the host excitation spectrum. The latter is generic for one-dimensional fluids; some two- and three-dimensional fluids also possess this structure. We expect that the impurity in any such fluid at zero temperature fails to thermalize.

We further apply the Boltzmann kinetic theory to describe the motion of an impurity under the action of a constant force. We find that the dynamical pattern crucially  depends on the impurity-to-host mass ratio $\eta$: A heavy impurity ($\eta>1$) experiences oscillations while the momentum of a light impurity ($\eta<1$) saturates without oscillations (see Fig. \ref{fig force}). The calculated damping rate of oscillations in the former case is found to be proportional to $F^2$, see Eq. \eqref{decrement}.  We thoroughly investigate the steady state of the impurity and find how it depends on the applied force. In particular, we find that at small forces the steady state momentum is proportional to $\sqrt{F}$ in the heavy impurity case (see Eq. \eqref{averageP1}) and is equal to $\eta \kF$ in the light impurity case (see Eq. \eqref{averageP2}).


The applicability conditions of our approach can be summarized as follows.  The impurity-host coupling $\gamma$ should be small and the mass ratio $\eta$ should be sufficiently far from unity, see Eq. \eqref{applicability conditions}. The latter condition excludes the vicinity of the integrable point $\eta=1$ from our consideration. In addition, the results for the impurity dynamics under the action of a force are valid when the force is larger than some threshold value. The latter is exponentially small in $\gamma$ for a heavy impurity, see Eq. \eqref{not too small force heavy}, and proportional to $\gamma^4$ for a light impurity, see Eq. \eqref{not too small force light}. The force is referred to as ``small'' when it is smaller than $\gamma^2 \kF/t_{\rm F}$ (but still larger than the threshold force).

The developed kinetic theory admits for a simple physical interpretation. The collision of the impurity with a host particle leads to the creation of elementary excitations of the fluid. The relaxation of the impurity momentum occurs through an uncorrelated chain of such events.
We have performed the calculations for a host gas in the Tonks-Girardeau limit, $\gamma_h=\infty$, and for a small impurity-host coupling $\gamma$. However, we expect that the qualitative picture of the impurity dynamics remains valid for a more general host with $\gamma_h<\infty$, as well as for moderate values of $\gamma$. Indeed, our main qualitative conclusions are based on robust kinematical considerations and on the classification of the host excitations which remain unchanged at any $\gamma_h$. The quantitative picture is of course modified by the renormalization of the scattering amplitudes in the Boltzmann equation and of spectral curves of the impurity and the host excitations, which depend on the values of the coupling constants.
This intuitive reasoning is further supported by the fact that under certain conditions the absence of the complete relaxation of the momentum of the impurity can be analytically demonstrated even for $\gamma\sim 1$ \cite{Lychkovskiy2013}.

The system we have studied is translation invariant. However our theory can be straightforwardly generalized to a smooth trapping potential, which is routinely used in experiments. Such generalization amounts to introducing a coordinate-dependent (and, if necessary, a time-dependent) force in the Boltzmann equation. It is possible to determine the dynamics of the impurity quantitatively by numerically solving such an equation.

One can also use the Boltzmann equation to investigate the dynamics of an impurity in a host gas at  finite temperature, which amounts to introducing thermal distribution function instead of step functions in transition rates \cite{CastroNeto1995}.
We expect that the presence of the thermal excitations in the host fluid will lift kinematical restrictions on the thermalization of the impurity.
Thus the impurity momentum in the absence of external force will eventually decay to zero with the decay rate strongly dependent on the temperature.
This effect can be used for local thermometry of the cold gas.

\acknowledgements{We thank E.~Burovski, M.~Zvonarev, M. Schecter and L. Glazman for illuminating discussions.
The present work was supported by the ERC grant 279738-NEDFOQ.
}


\appendix

\section{\label{appendix proof}}

Here we derive Eq. \eqref{integral eq for asymptotic distr} from Eq. \eqref{Boltzmann equation}.

The case $|p_0|<q_0$ is trivial since the r.h.s. of Eq. \eqref{Boltzmann equation} vanishes. Let us choose some $|p_0|>q_0$. The key idea is to consider $w_{p\ar k}$ for $|k|\leq q_0$ and for $q_0 <|k|\leq p_0$ separately. To emphasize this we  introduce special notations:
\be
\begin{array}{lll}
w_{p\ar k} = \chi_{p\ar k} & {\rm for } & q_0 <|k|\leq p_0;\\
w_{p\ar k} = \phi_{p\ar k} & {\rm for } & |k|\leq q_0.
\end{array}
\ee
(in both cases, however, $q_0 <|p|\leq p_0$)
The rational behind this separation is that the Boltzmann system for $\chi_{p\ar k}$ is closed:
\be\label{Boltzmann system above}
\dot{\chi}_{p\ar k}  = \sum_{q_0 <|k'|\leq p_0} \chi_{p\ar k'}
\underbrace{(-\delta_{k'k}\Gamma_k+\Gamma_{k'\ar k})}_{S_{k'\ar k}}.
\ee
Here we introduced a square $M \times M$ matrix $||S_{k'\ar k}||$, where $M \equiv2(p_0-q_0)/\delta k$. This matrix is triangular and its diagonal elements $S_{k\ar k}=-\Gamma_k$ are strictly negative. Two consequences  follow from these facts: $||S_{k'\ar k}||$ is invertible and
$
\chi_{p\ar k'}(\infty)=0.
$

The Boltzmann system for $\phi_{p\ar k}$, in contrast, is not closed:
\be
\dot{\phi}_{p\ar k}  = \sum_{q_0 <|k'|\leq p_0} \chi_{p\ar k'}
\Gamma_{k'\ar k}.
\ee
Integrating both sides over $t$ and taking into account Eq. \eqref{Boltzmann system above} and initial conditions one obtains
$$
\phi_{p\ar k}(\infty)= \int_0^\infty dt \sum_{q_0 <|k'|\leq p_0} \chi_{p\ar k'}(t)
\Gamma_{k'\ar k}
$$
$$
=\int_0^\infty dt
\mathop{\sum_{q_0 <|k'|\leq p_0}}_{q_0 <|k''|\leq p_0}
\dot{\chi}_{p\ar k''}(t)
\left(S^{-1}\right)_{k''\ar k'}\Gamma_{k'\ar k}
$$
\be
=-
\sum_{
q_0 <|k'|\leq p_0
}
\left(S^{-1}\right)_{p\ar k'}\Gamma_{k'\ar k}.
\ee
Contracting both sides with $S_{p_0p}$ one gets
\be
\sum_{q_0 <|p|\leq p_0}S_{p_0 p} \phi_{p\ar k}(\infty) = -\Gamma_{p\ar k},
\ee
which leads to the desired result \eqref{integral eq for asymptotic distr}.
%

\bibliography{Boltzmann}

\end{document}